\documentclass[prl,twocolumn,showpacs,superscriptaddress,preprintnumbers,amssymb,psfigs]{revtex4}
\usepackage{graphicx}
\usepackage{dcolumn}
\usepackage{bm}
\usepackage{wasysym}
\usepackage{txfonts}
\def\beq{\begin{equation}}
\def\eeq{\end{equation}}

\def\be{\begin{equation}}
\def\ee{\end{equation}}
\def\ba#1{\begin{array}{#1}}
\def\ea{\end{array}}
\def\bn{\begin{enumerate}}
\def\en{\end{enumerate}}

\begin{document}
\title{Susceptibility at the superfluid-insulator transition for one-dimensional disordered bosons}
\author{Shankar Iyer}
\author{David Pekker}
\author{Gil Refael}
\affiliation{Department of Physics, California Institute of
  Technology, MC 149-33, 1200 E.\ California Blvd., Pasadena, CA 91125}
\date{\today}
\begin{abstract}
A pair of recent Monte Carlo studies have reported evidence for and against a crossover from weak to strong-disorder criticality in the one-dimensional dirty boson problem. The Monte Carlo analyses rely on measurement of two observables: the effective Luttinger parameter $K_{\text{eff}}$ and the superfluid susceptibility $\chi$. The former quantity was previously calculated analytically, using the strong-disorder renormalization group (SDRG), by Altman, Kafri, Polkovnikov, and Refael. Here, we use an extension of the SDRG framework to find a non-universal anomalous dimension $\eta_{\text{sd}}$ characterizing the divergence of the  susceptibility with system size: $\chi \sim L^{2-\eta_{\text{sd}}}$. We show that $\eta_{\text{sd}}$ obeys the hyperscaling relation $\eta_{\text{sd}}=1/2K_{\text{eff}}$. We also identify an important obstacle to measuring this exponent on finite-size systems and comment on the implications for numerics and experiments. 
\end{abstract}
\maketitle


\indent Disordered bosonic systems pose theoretical challenges because of the unique pathologies of their non-interacting limits: at low
temperatures, bosons condense into a localized single-particle state, forming a configuration that is intrinsically unstable to interactions.  Therefore, Giamarchi and Schulz 
pioneered the study of the so-called ``dirty boson problem" by perturbing a strongly-interacting one-dimensional system with weak disorder.  They identified
a superfluid-insulator transition, belonging to the Kosterlitz-Thouless (KT) universality class, at which disorder is perturbatively irrelevant ~\cite{giamarchi1988anderson,ristivojevic2012phase}.
It was long believed that this universality \textit{always} characterizes the one-dimensional transition. In the past decade, the possibility has emerged that a novel criticality, 
also of KT type but with certain non-universal disorder-dependent features, takes over at sufficiently strong disorder strength.  
This ``strong-disorder criticality," first proposed by Altman, Kafri, Polkovnikov, and Refael ~\cite{altman2004pha,altman2008insulating,altman2010superfluid}, remains unconfirmed \cite{pollet2013classical}.
Recent Monte Carlo results by Hrahsheh and Vojta may provide evidence of the crossover between the two types of universality \cite{hrahsheh2012disordered}.  Meanwhile, experimental advances in 
various contexts, including cold atoms, spin systems, and dirty superconductors, have made it especially urgent to gain a better theoretical understanding
of the seemingly universal properties of the dirty boson problem ~\cite{allain2011gate,yu2011bose,schulte2005routes,white2009strongly,billy2008direct,roati2008anderson}.

\begin{figure}
\centering
\includegraphics[width=8cm]{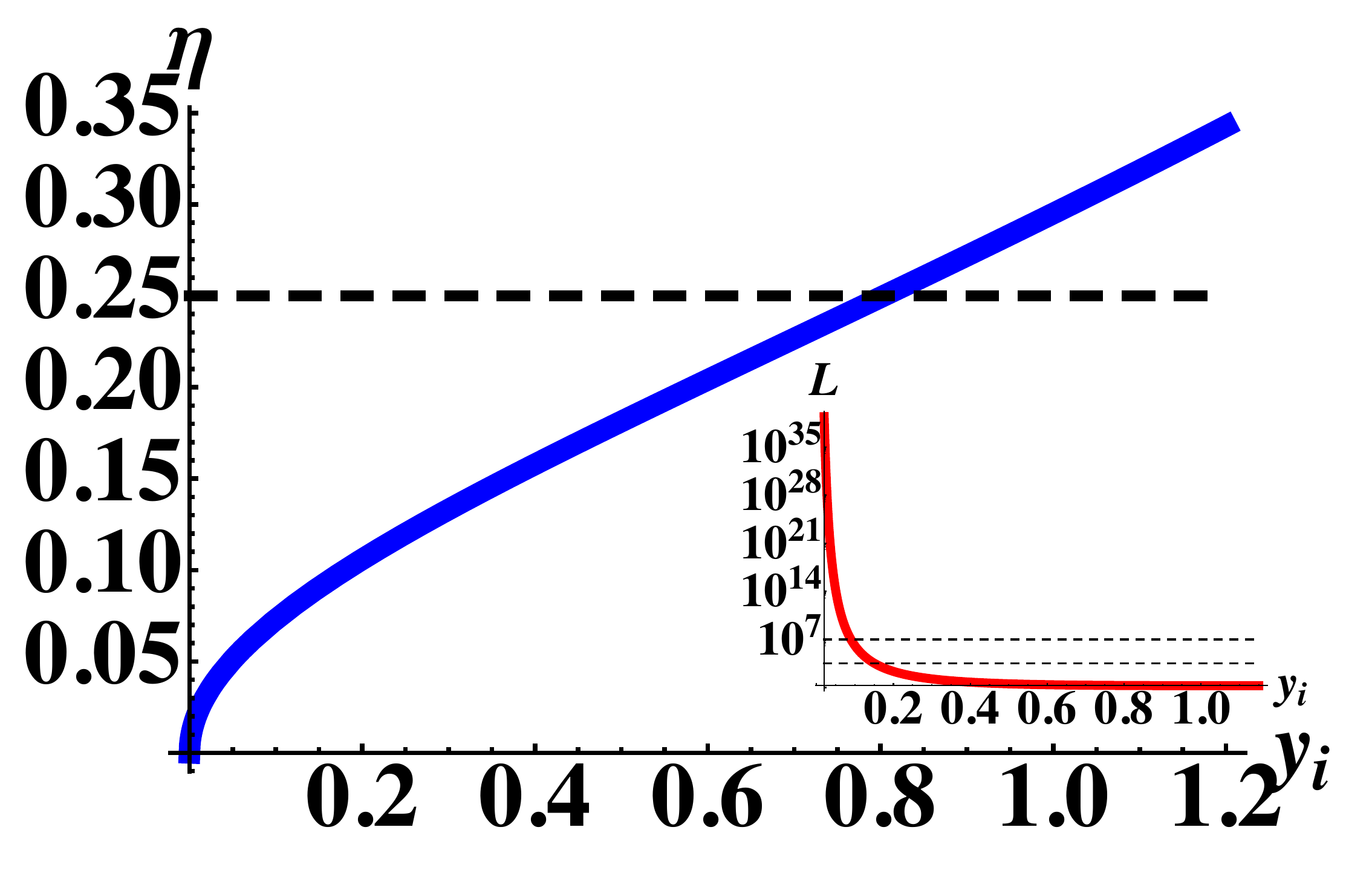}
\caption{In the main figure, the solid line shows the anomalous exponent $\eta_{\text{sd}}(y_i)$, as approximated in the small $y_i$ regime by (\ref{eq:anomalouseta}).
The dashed reference line shows the anomalous exponent of Giamarchi and Schulz \cite{giamarchi1988anderson}, and the crossing presumably indicates the crossover
from strong-disorder to weak-disorder criticality.  The inset shows the system size at which we can expect to cleanly observe the strong-disorder exponent $\eta_{\text{sd}}$.  
The two reference lines show, at increasing values of $L$, the typical system size probed by Monte Carlo \cite{hrahsheh2012disordered} and the system sizes available to numerical
SDRG.}
\label{fig:anomalouseta}
\end{figure}

\indent In this manuscript, we extend the analysis of the universal aspects of
the 1D  superfluid-insulator transition in the strong disorder regime.  In particular, we analytically calculate the superfluid susceptibility
near the transition. This affords us a new perspective on recent
numerical developments and allows us to clarify their relationship with
the theoretical SDRG framework. The model that we concentrate on is the particle-hole symmetric rotor model:
\begin{equation}
\label{eq:Hrot}
\hat{H}_{\text{rot}} = \sum^L_{j = 1} \left[U_j \hat{n}^2_j - J_j \cos{\left(\hat{\phi}_{j+1}-\hat{\phi}_j\right)}\right]
\end{equation}
This model can describe a 1D array of superconducting islands connected by Josephson junctions, and we assume
strong disorder in the on-site charging energies $U_j$ and Josephson couplings $J_j$.
Our principal result is that, at the strong-disorder transition, the divergence of the superfluid susceptibility is characterized
by an anomalous exponent:
\begin{equation}
\label{eq:chiexponent}
\lim_{L\rightarrow \infty} \frac{d \ln \chi}{d \ln L} = 2 - \eta_{\text{sd}}
\end{equation}
Here:
\begin{equation}
\label{eq:anomalouseta}
\eta_{\text{sd}} \approx \frac{1}{2\pi}\sqrt{2\left(e^{y_i}-1\right)}
\end{equation}
depends upon the bare disorder strength, parametrized by the quantity $y_i$.  We plot $\eta_{\text{sd}}$ as a function of $y_i$ in Figure \ref{fig:anomalouseta}.
The parameter $y_i$ can be understood if we imagine tuning the transition with the universal coupling distributions of the SDRG \cite{altman2004pha}:
$y_i = 0$ corresponds to a flat distribution of bare Josephson couplings, and as $y_i$ increases, the bare Josephson coupling
distribution becomes progressively more strongly peaked near the RG scale, effectively reducing the disorder strength.  Thus, 
the anomalous dimension monotonically increases as the disorder strength decreases, and the weak disorder universality presumably takes
over when $\eta_{\text{sd}} \approx \frac{1}{4}$, the value at the Giamarchi-Schulz transition \cite{giamarchi1988anderson}.  Throughout
the strong-disorder regime, our prediction (\ref{eq:anomalouseta}) approximately obeys:
\begin{equation}
\label{eq:hyperscaling}
\eta_{\text{sd}} = \frac{1}{2K_{\text{eff}}}
\end{equation}
where $K_{\text{eff}}$ is the Luttinger parameter predicted by Altman et al. \cite{altman2010superfluid}.  This scaling relation 
follows in clean systems from a Kubo formula for the susceptibility, and Monte Carlo results suggest that it may be valid in
strongly disordered systems as well \cite{hrahsheh2012disordered}.


\indent Below, we describe the calculation leading to our exponent $\eta_{\text{sd}}$.  We begin by making some 
additional comments on our model (\ref{eq:Hrot}).  We then proceed to briefly outline the strong-disorder renormalization group (SDRG) procedure.  
Next, we qualitatively describe and then perform the calculation, comparing intermediate analytical predictions 
to a numerical implementation of the SDRG when possible.  Finally, we comment on implications of our results for 
theory, numerics, and experiments.
  
\indent The rotor model (\ref{eq:Hrot}) can be viewed as a disordered Bose-Hubbard model at large commensurate filling.  However, it strictly omits diagonal (i.e., chemical
potential) disorder and, therefore, exhibits an exact particle-hole symmetry ($\hat{n}_j \rightarrow -\hat{n}_j$, $\hat{\phi}_j \rightarrow -\hat{\phi}_j$).  
Altman et al. showed that this symmetry results in an incompressible Mott glass phase intervening between the superfluid and Mott insulating phases in the phase diagram ~\cite{altman2008insulating,weichman2008particle}.  
The Mott glass replaces the Bose glass phase of the generic dirty boson problem \cite{fisher1988onset,fisher1989boson}.  Despite this, the universal properties of 
the strong-disorder transition do \textit{not} depend on the special symmetry properties of the model, and we perform our calculation in the particle-hole symmetric
model for analytical convenience.

\begin{figure} \centering
\begin{minipage}[b]{0.4cm}
       {\bf (a)}

       \vspace{1cm}
\end{minipage}
\begin{minipage}[t]{7.9cm}
       \includegraphics[width=7.8cm]{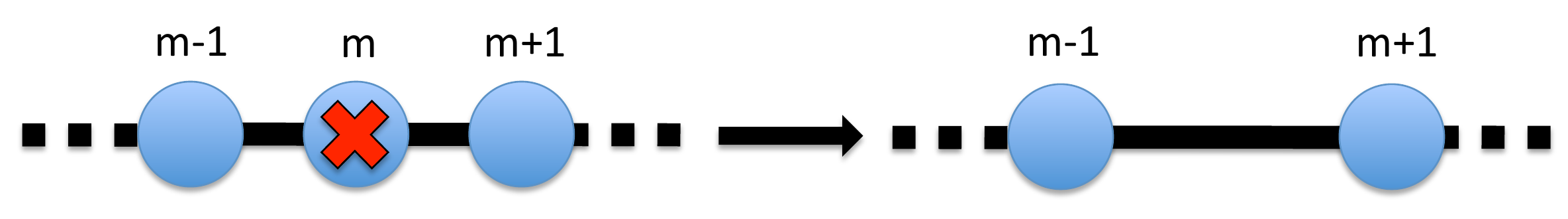}
\end{minipage}\\
\begin{minipage}[b]{0.4cm}
       {\bf (b)}

       \vspace{1cm}
\end{minipage}
\begin{minipage}[t]{7.9cm}
       \includegraphics[width=7.8cm]{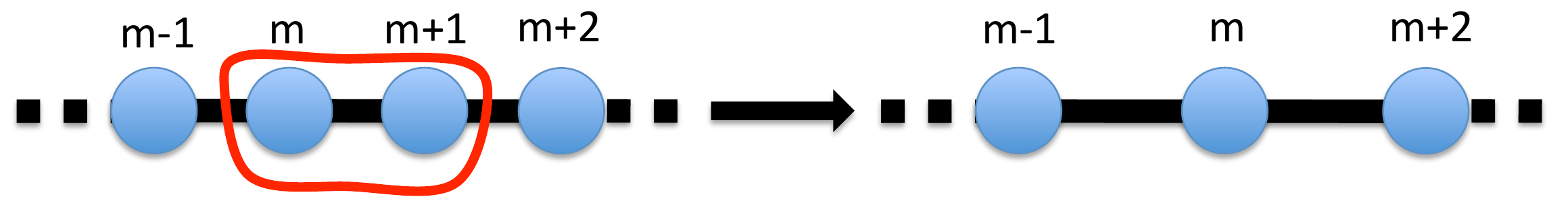}
\end{minipage}\\

\caption{In panel (a), the site decimation RG step, and in panel (b), the link decimation RG step.}
\label{fig:RGsteps}
\end{figure}

\indent We perform our calculation using the SDRG ~\cite{dasgupta1980low,bhatt1982scaling,fisher1994random,fisher1995critical}.
We iteratively find the strongest coupling in the problem (which sets the \textit{RG scale} $\Omega$) and locally choose the ground state 
to satisfy that term in the Hamiltonian.   We then account for the effect of neighboring couplings perturbatively.  This process gradually lowers $\Omega$, 
leading to a low-energy description of the system.  For the rotor model (\ref{eq:Hrot}), there are two possible RG steps.  If the dominant coupling is
a charging energy (i.e., $\Omega = U_m$),  then it is unlikely that the particle number will fluctuate strongly on site $m$.  To zeroth order, we can set $n_m = 0$ and 
calculate perturbative corrections from the Josephson couplings penetrating this site.  This \textit{site-decimation} procedure, pictured in Figure \ref{fig:RGsteps}.a, 
leads to an effective coupling between sites $m-1$ and $m+1$:
\begin{equation}
\label{eq:sitedec}
\frac{\tilde{J}_{m-1}}{\Omega} = \frac{J_{m-1}J_m}{\Omega^2}
\end{equation}
Instead, the RG scale may be set by a Josephson coupling (i.e., $\Omega = J_m$).  Then, it is unlikely that there will
be phase slips between sites $m$ and $m+1$, and we may merge these two sites into one, yielding an effective charging energy for a new cluster site:
\begin{equation}
\label{eq:linkdec}
\frac{\Omega}{\tilde{U}_m} = \frac{\Omega}{U_m}+\frac{\Omega}{U_{m+1}}
\end{equation}
This process of link decimation is shown in Figure \ref{fig:RGsteps}.b \cite{altman2004pha}.

\indent Altman et al.\ wrote integrodifferential equations describing the flow of the distributions of the variables $\zeta_j = \frac{\Omega}{U_j}-1$ and $\beta_j = \ln{\left(\frac{\Omega}{J_j}\right)}$ as a function of the \textit{RG time}
$\Gamma = \ln{\left(\frac{\Omega_i}{\Omega}\right)}$, where $\Omega_i$ is the initial RG scale.  We direct the reader to Reference \cite{altman2004pha}
for these flow equations and here only quote the solution forms for the distributions:
\begin{equation}
\label{eq:universalforms}
f(\zeta,\Gamma) = f_0 e^{-f_0\zeta}, \text{  } g(\zeta,\Gamma) = g_0 e^{-g_0\beta}
\end{equation}
where $f_0 = f(0,\Gamma)$ and $g_0 = g(0,\Gamma)$ satisfy:
\begin{equation}
\label{eq:redfloweq}
\frac{df_0}{d\Gamma} = f_0(1-g_0), \text{  } \frac{dg_0}{d\Gamma} = -f_0 g_0
\end{equation}
The quantity $\epsilon = f_0-g_0+\ln g_0+1$ is an invariant and natural tuning parameter of the flows (\ref{eq:redfloweq}).  
When $\epsilon = 0$, the flow terminates at an unstable fixed point at $(f_0,g_0) = (0,1)$ 
that controls the transition between insulator ($\epsilon > 0$) and superfluid ($\epsilon < 0$).  This fixed point corresponds
to a \textit{classical} model with vanishing charging energies.  Expanding about this point by defining $g_0(\Gamma) = 1+y(\Gamma)$, we
can approximate the critical flows:
\begin{equation}
\label{eq:criticalflows}
f_0(\Gamma) \approx \frac{2}{\gamma^2}, \text{  } y(\Gamma) \approx \frac{2}{\gamma}
\end{equation}
Here, $y_i \equiv y(0)$ and:
\begin{equation}
\label{eq:Gammetilde}
\gamma(\Gamma) \equiv \Gamma + \frac{2}{y_i}
\end{equation}


\indent Consider perturbing the rotor model (\ref{eq:Hrot}) with an ordering field:
\begin{equation}
\label{eq:perturbation}
\hat{H}' = -h \sum_j \cos{(\hat{\phi}_j)}
\end{equation}
The superfluid susceptibility is the linear response:
\begin{equation}
\label{eq:susc}
\chi = \frac{1}{L} \sum_j  \frac{\partial \langle \cos(\hat{\phi}_j) \rangle}{\partial h}|_{h = 0}
\end{equation}
The RG builds clusters of various sizes (through link decimation steps) and then removes these clusters 
from the chain (through site decimation steps).  Each cluster is an approximately
independent superfluid island, and our goal is to accumulate the contributions to $\chi$ from all such islands.  
We do this by following the RG to a time $\Gamma_f$ at which the initial $L$-site chain has been renormalized to a single cluster.  Then,
we take into account the contribution of the final cluster separately:
\begin{equation}
\label{eq:susc2}
\chi = \frac{X_f}{L} + \int_0^{\Gamma_f} d\Gamma \rho(\Gamma) X_{\text{clust}}(\Gamma)
\end{equation}
Here, $X_{\text{clust}}(\Gamma)$ is the \textit{extensive} superfluid susceptibility of clusters decimated at time $\Gamma$,
$\rho(\Gamma)$ is the density (number per unit area) of these clusters, and $X_f$ is the extensive susceptibility of the final
cluster.

\indent In implementing the calculation (\ref{eq:susc2}), we also want to keep in mind that, in one dimension, 
true long-range order does not exist within superfluid clusters; therefore, we must account 
for the internal fluctuations that the SDRG neglects.  The link decimation procedure implies absence of phase slips within the cluster,
and this makes the effective cluster Hamiltonian quadratic.  A uniform, quadratic Hamiltonian is easy to 
study analytically, so we adopt a uniformization procedure (described below) to calculate the cluster susceptibilities.
Uniformization is reasonable as long as the susceptibility of a disordered, quadratic chain self averages.  Numerical checks
on moderately sized ($L = 100$) systems seem to indicate that this is the case; nevertheless, the uniformization procedure
is an uncontrolled approximation whose ultimate justification is consistency with the Monte Carlo results \cite{hrahsheh2012disordered}.


\indent We now proceed to the details of the calculation.  We first use our solutions (\ref{eq:criticalflows}) to calculate 
the number of sites remaining in the system at RG time $\Gamma$ (see equation (8) of Reference \cite{altman2004pha}):
\begin{equation}
\label{eq:Ngamma}
N(\Gamma) \approx L\frac{4e^{\frac{2}{y_i}}}{y^2_i} \frac{e^{-\gamma(\Gamma)}}{\gamma(\Gamma)^2}
\end{equation}
Reasoning that a fraction $f_0(\Gamma)d\Gamma$ of these get site-decimated in the interval $(\Gamma,\Gamma+d\Gamma)$, 
we find:
\begin{equation}
\label{eq:rhogamma}
\rho(\Gamma)d\Gamma \approx \frac{8e^{\frac{2}{y_i}}}{y^2_i} \frac{e^{-\gamma(\Gamma)}}{\gamma(\Gamma)^4} d\Gamma
\end{equation}
Also, using equation (\ref{eq:Ngamma}), we can infer the renormalization time $\Gamma_f$ by setting $N(\Gamma_f) = 1$.  
This yields an expression that can be iteratively inverted to yield:
\begin{equation}
\label{eq:gammaf}
\Gamma_f \approx \ln{\left(\frac{4e^{\frac{2}{y_i}}}{y_i^2}L\right)} - 2\ln{\ln{\left(\frac{4e^{\frac{2}{y_i}}}{y_i^2}L\right)}} - \frac{2}{y_i}
\end{equation}

\begin{figure}
\centering
\includegraphics[width=8cm]{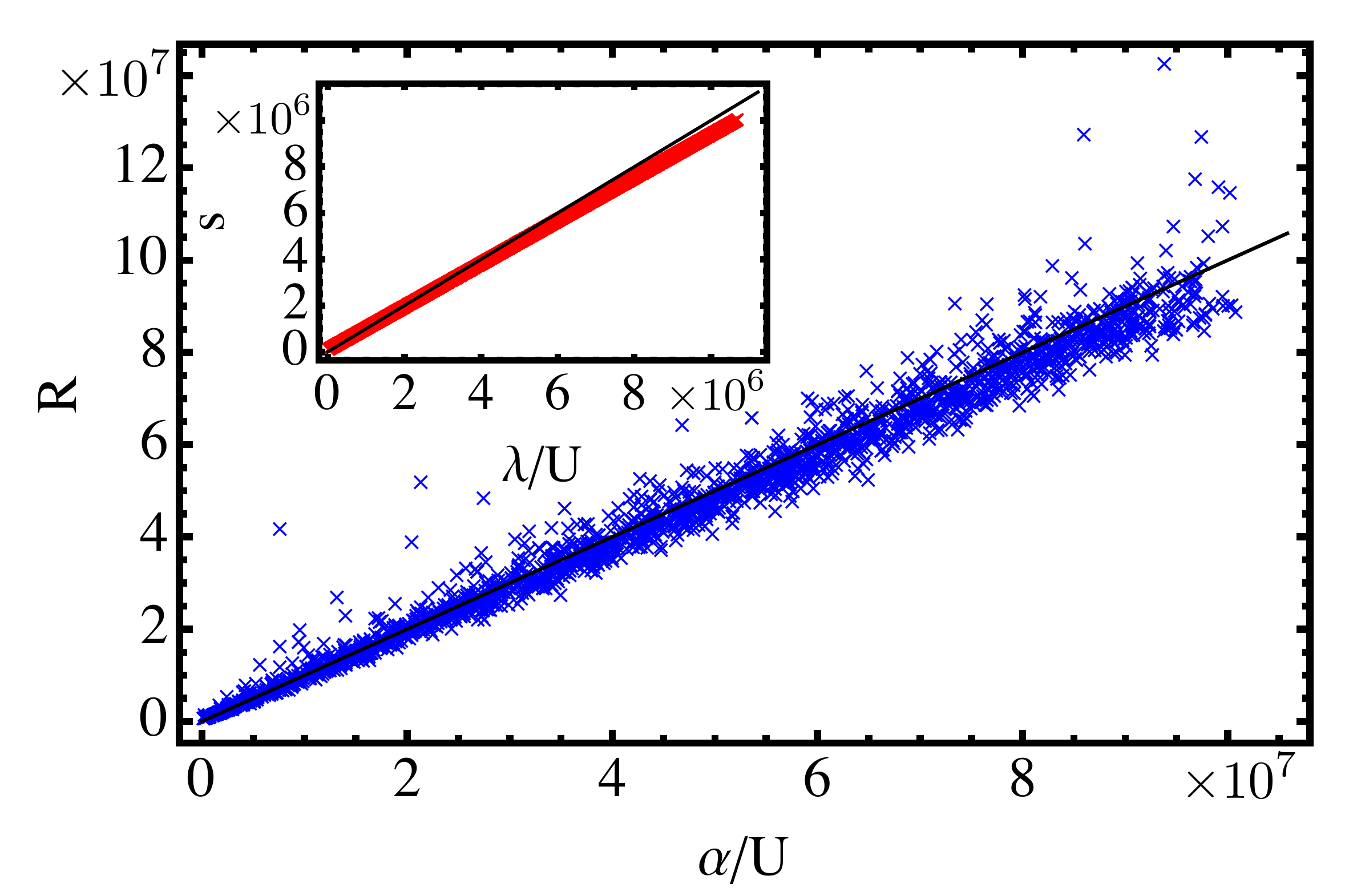}
\caption{In the main figure, a check of the result (\ref{eq:Rbarsolution}) for $\bar{R}(\zeta,\Gamma)$.  Note that $\alpha = \left[\exp{\left(y_i-\frac{2}{\gamma(\Gamma)}\right)}-1\right]$.
We stop the numerical RG for $y_i = 0.1$ when only two sites are remaining and pool $10^3$ samples.  In the inset, a similar check of the result (\ref{eq:sbar}) for $\bar{s}(\zeta,\Gamma)$.
Here, $\lambda = \frac{\Omega_i y_i^2}{2}$.  The reference lines show the analytical predictions, and the inset blocks one outlier of the main figure.}
\label{fig:Rcheck}
\end{figure}

\indent We also need to compute certain ``internal" properties of the clusters decimated at time $\Gamma$, including their typical size $\bar{s}$ (i.e., the 
number of bare sites that they represent) and statistical properties of the internal couplings.  In our cluster uniformization procedure, we
will define the effective uniform charging energies and Josephson couplings as the averages:
\begin{equation}
\label{eq:uniformizedcouplings}
\frac{1}{U_{\text{unif}}} \equiv \frac{1}{s}\sum_{j \in c} \frac{1}{U_j} \text{, } \frac{1}{J_{\text{unif}}} \equiv \frac{1}{s}\sum_{j \in c} \frac{1}{J_j}
\end{equation}
where the sums are taken over all the sites or links of which the cluster is built.  While $U_{\text{unif}} = s\Omega$ for clusters at the RG scale,
finding $J_{\text{unif}}$ requires an extension of the RG to keep track of the distribution $\tilde{f}(\zeta,R,\Gamma)$
where $R$ is the sum:
\begin{equation}
\label{eq:Rdefine}
R = \sum_{j \in c} \frac{1}{J_j}
\end{equation}
Such an extension was described in Reference \cite{altman2010superfluid}.  Here, we are interested in the average value of $R$ as a function of $\zeta$ and $\Gamma$:
\begin{equation}
\label{eq:ourRavg}
\bar{R}(\zeta,\Gamma) = \frac{\int dR R \tilde{f}(\zeta,R,\Gamma)}{\int dR \tilde{f}(\zeta,R,\Gamma)}
\end{equation}
We can formulate an equation governing the evolution of this average near criticality:
\begin{eqnarray}
\label{eq:Rfloweq}
\frac{\partial \bar{R}}{\partial \Gamma} & = &  (1+\zeta)\frac{\partial \bar{R}}{\partial \zeta} \nonumber \\
                                              & \quad & + \bar{R}\left[f_0(\Gamma)+1-g_0(\Gamma)+(1+\zeta)\frac{\partial \ln{f}}{\partial \zeta}-\frac{\partial \ln{f}}{\partial \Gamma}\right] \nonumber \\
                                              & \quad & + 2f_0(\Gamma)g_0(\Gamma)\int_0^\zeta d\zeta' \bar{R}(\zeta',\Gamma) + \zeta\frac{f_0(\Gamma)g_0(\Gamma)}{\Omega} 
\end{eqnarray}
In the large $\Gamma$ limit, this equation can be solved when we insert the solutions (\ref{eq:universalforms}) and (\ref{eq:criticalflows}):
\begin{equation}
\label{eq:Rbarsolution}
\bar{R}(\zeta,\Gamma) \approx \frac{e^\Gamma}{\Omega_i} (1+\zeta)\left[\exp{\left(y_i-\frac{2}{\gamma(\Gamma)}\right)}-1\right]
\end{equation}
Proceeding along similar lines, we can find the mean cluster size $\bar{s}(\zeta,\Gamma)$:
\begin{equation}
\label{eq:sbar}
\bar{s}(\zeta,\Gamma) \approx \frac{y^2_i}{2} (1+\zeta) e^\Gamma
\end{equation}

\indent We now pause to compare the predictions (\ref{eq:Rbarsolution}) and (\ref{eq:sbar}) to numerical RG.  We run the RG on $L = 10^7$ lattices, beginning with the attractor
distributions (\ref{eq:universalforms}), for various values of $y_i$.  We periodically interrupt the procedure and check if the numerically
generated distributions of $\bar{R}$ and $\bar{s}$ match the analytical expectations.  In Figure \ref{fig:Rcheck}, we plot results
for $y_i = 0.1$, showing that the predictions remain valid until the effective renormalized chain consists of only two sites (i.e., after $10^7 -2$ RG steps).

\indent In its final two steps, the numerical RG typically merges the two remaining sites and then site-decimates the resulting cluster.  
At this stage, the comparison of analytics and numerics is subtle.
Essentially, it is illegitimate to estimate $\Gamma$ based on the final site's charging energy, since there is no longer any distribution of sites
or couplings.  Instead, we can assume that the RG time is $\Gamma_f$ and rewrite equation (\ref{eq:Rbarsolution}) to eliminate $\zeta$ in favor of the effective charging energy $U_f$.
The numerical results then confirm the following relationship for $\bar{R}_f$:
\begin{equation}
\label{eq:UeffRbar}
U_f \bar{R}_f \approx \exp{\left(y_i-\frac{2}{\gamma_f}\right)}-1
\end{equation}
where $\gamma_f = \gamma(\Gamma_f)$.
We can also circumvent equation (\ref{eq:sbar}) and compute the size of the final cluster by finding the fraction of the chain that has been site-decimated by time $\Gamma_f$:
\begin{equation}
\label{eq:fdec}
\int_0^{\Gamma_f} d\Gamma \rho(\Gamma) \bar{s}(0,\Gamma) \approx \frac{1}{2}y_i^2
\end{equation}
Thus, the final cluster represents approximately $(1-\frac{1}{2}y_i^2)L$ sites of the original chain.  

\indent The final ingredient that we need is the calculation of the susceptibility of the uniformized cluster.  We begin with the
uniformized quadratic Hamiltonian:
\begin{equation}
\label{eq:Hcluster}
H_{\text{clust}} = \frac{J_{\text{unif}}}{2}\sum_{j=1}^{\bar{s}} (\hat{\phi}_j-\hat{\phi}_{j+1})^2 + U_{\text{unif}}\sum_{j=1}^{\bar{s}} \hat{n}^2_j
\end{equation}
and use standard path integral techniques.  Here, we simply quote the result:
\begin{equation}
\label{eq:chiclustresult}
X_{\text{clust}} \approx  \frac{(\pi e^{\gamma_E})^{-\eta} \bar{s}^{3-\eta}}{U_{\text{unif}}} \left[1 + \eta\ln \eta+\eta\ln{\left(\pi e^{\gamma_E}\right)}\right]
\end{equation}
with $\gamma_E \approx 0.5772$ is the Euler gamma and :
\begin{equation}
\label{eq:etadefinition}
\eta \equiv \frac{1}{2\pi} \sqrt{\frac{2U_{\text{unif}}}{J_{\text{unif}}}} = \frac{1}{2\pi} \sqrt{2UR}
\end{equation}
Note that $U = \bar{s}U_{\text{unif}}$.

\indent We can now assemble the final result.  Equations (\ref{eq:chiclustresult}) and (\ref{eq:UeffRbar}) allow us to compute the final cluster's contribution to the susceptibility:
\begin{eqnarray}
\label{eq:finalclusterchi}
\chi_f  & \approx & \frac{2}{y_i^2 \Omega_i}(\pi e^{\gamma_E})^{-\eta(\gamma_f)} \left(1-\frac{y^2_i}{2}\right)^{3-\eta(\gamma_f)}L^{2-\eta(\gamma_f)} \nonumber \\
            & \quad & \times [1+\eta(\gamma_f)\ln(\pi e^{\gamma_E}) + \eta(\gamma_f) \ln{\eta(\gamma_f)}]
\end{eqnarray}
where:
\begin{eqnarray}
\label{eq:etagamma}
\eta(\gamma) = \frac{1}{2\pi}\sqrt{2\left[\exp{\left(y_i-\frac{2}{\gamma(\Gamma)}\right)}-1\right]}
\end{eqnarray}
On the other hand, determining the contribution of the subleading clusters involves evaluating the integral:
\begin{eqnarray}
\label{eq:subleadingchi}
\chi_{\text{sl}} & \approx & \frac{2y_i^2}{\Omega_i}  \int_{\frac{2}{y_i}}^{\gamma(\Gamma_f)} d\gamma \left(\frac{\pi y_i^2}{2} e^{\gamma_E}e^{-\frac{2}{y_i}}\right)^{-\eta(\gamma)} \frac{e^{(2-\eta(\gamma))\gamma}}{\gamma^4} \nonumber \\
                         & \quad &   \quad \quad \times [1+\eta(\gamma)\ln(\pi e^{\gamma_E}) + \eta(\gamma)\ln{\eta(\gamma)}]
\end{eqnarray}
At small $y_i$, the density of subleading clusters is strongly suppressed, and these clusters have little opportunity to contribute to 
the susceptibility. 
In any case, in the thermodynamic limit, both contributions (\ref{eq:finalclusterchi}) and (\ref{eq:subleadingchi}) diverge as $L^{2-\eta_{\text{sd}}}$,
where $\eta_{\text{sd}} = \lim_{\gamma \rightarrow \infty} \eta(\gamma)$ is the anomalous exponent (\ref{eq:anomalouseta}).
Thus, $\eta_{\text{sd}}$ is the principal result of our work.

\indent We must, however, question when this anomalous exponent can be observed.  Equation (\ref{eq:etagamma}) reveals
that the thermodynamic limit requires:
\begin{equation}
\label{eq:tlcondition}
\Gamma_f \gg \frac{2}{y_i}
\end{equation}
This implies:
\begin{equation}
\label{eq:tlcondition2}
L \gg \frac{y_i^2}{4}\exp{\left(\frac{2}{y_i}\right)}
\end{equation}
For small values of $y_i$, this corresponds to an enormous length scale that is inaccessible to Monte Carlo and even 
numerical SDRG.  We plot this length scale as a function of $y_i$ in the inset of Figure \ref{fig:anomalouseta}.  
This indicates the need for caution in interpreting the results of Hrahsheh and Vojta \cite{hrahsheh2012disordered}: due to strong finite-size effects,
their measurement of the anomalous dimension likely underestimates the ``true" thermodynamic value. 
Here, an underestimate actually moves $\eta$ further from the value at the Giamarchi-Schulz transition, $\eta = \frac{1}{4}$, and closer to the scaling
result, $\eta = 0$.  Since various laboratory systems (e.g., ultracold atoms) reach only moderate values of $L$, this obstacle to cleanly observing 
$\eta_{\text{sd}}$ could be experimentally relevant.

\indent Following the Monte Carlo results of Hrahsheh and Vojta, Pollet et al.\ argued that the one-dimensional dirty boson problem is characterized by a slow classical renormalization of Josephson couplings, 
beyond which the criticality of Giamarchi and Schulz sets in at inaccessibly large length scales \cite{pollet2013classical}.  
In the strong-disorder scenario, late stages of the critical RG flow are also dominated by link decimations; however, 
rare site decimations have a dramatic effect in renormalizing the Josephson coupling distribution, and this has crucial consequences for the susceptibility.  
Thus, an exceedingly slow and apparently classical renormalization flow can also be a precursor of strong-disorder criticality.

\indent As a final point, we reiterate that our anomalous exponent (\ref{eq:anomalouseta}) approximately obeys the scaling relation (\ref{eq:hyperscaling}), which usually
follows from a  Kubo formula in clean systems.  In our uniformization procedure, we essentially calculate $K$ for each cluster and then integrate the resulting correlation function to find the cluster's susceptibility: 
hence, we build in the Kubo formula for each cluster.  More surprisingly, $\eta = \frac{1}{2K}$ was observed in quantum Monte Carlo \cite{hrahsheh2012disordered}.  One of the virtues of our methodology is that it explains why
the scaling relation is obeyed on finite-size lattices, even as $\eta$ and $K$ slowly drift to their thermodynamic values. 

\indent We thank T. Vojta for sharing his numerical results with us at the 2012 workshop on Quantum Matter from the Nano- to the Macroscale at the MPIPKS-Dresden.  We also acknowledge T. Giamarchi and L. Pollet for helpful conversations. We are grateful to the MPIPKS-Dresden, the KITP, and the Aspen Center for Physics for their hospitality and acknowledge financial support from the IQIM, an NSF center supported in part by the Moore foundation.  Additionally, DP and GR are grateful for financial support from the Lee A. DuBridge Fellowship and Packard foundation respectively.

\bibliographystyle{unsrt}
\bibliography{allrefs}

\begin{thebibliography}{10}

\bibitem{giamarchi1988anderson}
T.~Giamarchi and H.J. Schulz.
\newblock {\em Physical Review B}, 37(1):325, 1988.

\bibitem{ristivojevic2012phase}
Z.~Ristivojevic, A.~Petkovi{\'c}, P.~Le~Doussal, and T.~Giamarchi.
\newblock {\em Physical Review Letters}, 109(2):026402, 2012.

\bibitem{altman2004pha}
E.~Altman, Y.~Kafri, A.~Polkovnikov, and G.~Refael.
\newblock {\em Physical Review Letters}, 93(15):150402, 2004.

\bibitem{altman2008insulating}
E.~Altman, Y.~Kafri, A.~Polkovnikov, and G.~Refael.
\newblock {\em Physical Review Letters}, 100(17):170402, 2008.

\bibitem{altman2010superfluid}
E.~Altman, Y.~Kafri, A.~Polkovnikov, and G.~Refael.
\newblock {\em Physical Review B}, 81(17):174528, 2010.

\bibitem{pollet2013classical}
L.~Pollet, N.V. Prokof'ev, and B.V. Svistunov.
\newblock {\em Physical Review B}, 87:144203, 2013.

\bibitem{hrahsheh2012disordered}
F.~Hrahsheh and T.~Vojta.
\newblock {\em Physical Review Letters}, 109(26):265303, 2012.

\bibitem{allain2011gate}
A.~Allain, Z.~Han, and V.~Bouchiat.
\newblock {\em Nature Materials}, 11(7):590, 2012.

\bibitem{yu2011bose}
R.~Yu, L.~Yin, N.S Sullivan, J.S. Xia, C.~Huan, A.~Paduan-Filho, N.F.
  Oliveira~Jr, S.~Haas, A.~Steppke, C.F. Miclea, et~al.
\newblock {\em Nature}, 489(7416):379, 2012.

\bibitem{schulte2005routes}
T.~Schulte, S.~Drenkelforth, J.~Kruse, W.~Ertmer, J.~Arlt, K.~Sacha,
  J.~Zakrzewski, and M.~Lewenstein.
\newblock {\em Physical Review Letters}, 95(17):170411, 2005.

\bibitem{white2009strongly}
M.~White, M.~Pasienski, D.~McKay, S.Q. Zhou, D.~Ceperley, and B.~DeMarco.
\newblock {\em Physical Review Letters}, 102(5):55301, 2009.

\bibitem{billy2008direct}
J.~Billy, V.~Josse, Z.~Zuo, A.~Bernard, B.~Hambrecht, P.~Lugan, D.~Cl{\'e}ment,
  L.~Sanchez-Palencia, P.~Bouyer, and A.~Aspect.
\newblock {\em Nature}, 453(7197):891, 2008.

\bibitem{roati2008anderson}
G.~Roati, C.~DÕErrico, L.~Fallani, M.~Fattori, C.~Fort, M.~Zaccanti,
  G.~Modugno, M.~Modugno, and M.~Inguscio.
\newblock {\em Nature}, 453(7197):895, 2008.

\bibitem{weichman2008particle}
P.B. Weichman and R.~Mukhopadhyay.
\newblock {\em Physical Review B}, 77(21):214516, 2008.

\bibitem{fisher1988onset}
D.S. Fisher and M.P.A. Fisher.
\newblock {\em Physical Review Letters}, 61(16):1847--1850, 1988.

\bibitem{fisher1989boson}
M.P.A. Fisher, P.B. Weichman, G.~Grinstein, and D.S. Fisher.
\newblock {\em Physical Review B}, 40(1):546, 1989.

\bibitem{dasgupta1980low}
C.~Dasgupta and S.~Ma.
\newblock {\em Physical Review B}, 22(3):1305, 1980.

\bibitem{bhatt1982scaling}
RN~Bhatt and PA~Lee.
\newblock {\em Physical Review Letters}, 48(5):344, 1982.

\bibitem{fisher1994random}
D.S. Fisher.
\newblock {\em Physical Review B}, 50(6):3799, 1994.

\bibitem{fisher1995critical}
D.S. Fisher.
\newblock {\em Physical Review B}, 51(10):6411, 1995.

\end{thebibliography}
\end{document}